\documentclass[aps,prd,twocolumn,showpacs,floats,floatfix,letterpaper,nofootinbib,superscriptaddress,]{revtex4-1}

\usepackage{amssymb,amsmath,latexsym,mathrsfs}
\usepackage{graphicx,subfigure}
\usepackage{epsfig}
\usepackage{varioref,xr-hyper}
\usepackage{color}
\usepackage{multirow}
\usepackage{array}

\newcommand{\neff}{N_{\rm {eff}}}
\newcommand{\cvis}{c_{\rm vis}^2}
\newcommand{\ceff}{c_{\rm eff}^2}

\begin{document}


\title{Dark Radiation in extended cosmological scenarios}

\author{Maria Archidiacono}
\affiliation{Physics Department and INFN, Universita' di Roma 
	``La Sapienza'', Ple.\ Aldo Moro 2, 00185, Rome, Italy}
\author{Elena Giusarma}
\affiliation{IFIC, Universidad de Valencia-CSIC, 46071, Valencia, Spain}
\author{Alessandro Melchiorri$^1$}
\author{Olga Mena$^2$}

\begin{abstract}
Recent cosmological data have provided evidence for a "dark" relativistic background at high
statistical significance. Parameterized in terms of the number of relativistic degrees of freedom $\neff$,
however, the current data seems to indicate a higher value than the one expected in the standard
scenario based on three active neutrinos. This dark radiation component can be characterized not only by its abundance but also by its clustering properties, as its effective sound speed and its viscosity parameter. It is therefore crucial to study the correlations among the dark radiation properties and key cosmological parameters, as the dark energy equation of state or the running of the scalar spectral index, with current and future CMB data. We find that dark radiation with viscosity parameters different from their standard values may be misinterpreted as an evolving dark energy component or as a running spectral index in the power spectrum of primordial fluctuations.

\end{abstract}


\maketitle

\section{Introduction}

The cosmological abundance of relativistic and "dark" particles as active or sterile neutrinos~\footnote{There is no fundamental symmetry in nature forcing  a definite number of right-handed (sterile) neutrino species, as those are allowed in the Standard Model fermion content.} or any other light particle (as axions, for instance) is usually parameterized in terms of the number of relativistic degrees of freedom $\neff$, being $\neff=3.046$ the standard value for three (active) neutrino species. 

Several recent cosmological data analyses have constrained the value of $\neff$ with increasing accuracy (see, for instance, Refs.~\cite{Mangano:2006ur,Hamann:2007pi,Reid:2009nq,Komatsu:2010fb,Hamann:2010pw,Hamann:2011hu,Nollett:2011aa,Giusarma:2011zq}) providing a clear evidence for a dark relativistic background at high statistical
significance. The latest measurements of Cosmic Microwave Background (CMB) anisotropies at arc-minute angular scales from the South Pole Telescope (SPT)~\cite{spt} and the Atacama Cosmology Telescope (ACT)~\cite{act}, when combined with other cosmological data sets,
yield the constraint $\neff=4.08_{-0.68}^{+0.71}$ at $95 \%$ confidence level (CL)~\cite{darkr} (see also \cite{Hou:2011ec} and \cite{Smith:2011es}), showing evidence
for dark radiation (i.e. $\neff>0$) at more than seven standard deviations, suggesting values higher than those
expected in the standard scenario.
This hint, that clearly must be tested by future CMB measurements as those expected, for example, by the Planck
satellite \cite{Planck:2006aa}, is interesting since an extra relativistic component in the standard three active neutrino model could be explained by a sterile neutrino. Models with one additional $\sim 1$~eV massive sterile neutrino, i.e. the so-called (3+1) models, were introduced to explain LSND short baseline antineutrino data~\cite{Aguilar:2001ty} by means of neutrino oscillations.  Up to date cosmological constraints on the number of massive sterile neutrino species have been presented in Refs.~\cite{Melchiorri:2008gq,rt,us,Giusarma:2011zq}. However, a larger value for $\neff$ could also arise from different physics, related to axions~\cite{axions}, decay of non-relativistic matter~\cite{decay}, gravity waves~\cite{gw}, extra dimensions~\cite{extra}, dark energy~\cite{ede} or asymmetric dark matter models~\cite{Blennow:2012de}. The constraints on $\neff$ are also affected by possible interactions of the dark radiation component with the dark matter component, see Refs.~\cite{Mangano:2006mp,Serra:2009uu,Blennow:2012de}.

Information on the dark relativistic background, however, can be obtained not only from its effects on the expansion rate of the universe but also from its clustering properties. For example, if dark radiation is made of massless neutrinos it should behave as relativistic particles also from the point of view of perturbation theory. Following the definition introduced by Hu~\cite{hudm}, this means that the dark radiation component should have an effective sound speed $\ceff$ and a viscosity parameter $\cvis$ such that $\ceff=\cvis=1/3$.  
These perturbation parameters can be constrained through measurements of the CMB anisotropies since dark radiation is coupled trough gravity with all the remaining components~\cite{hudm2}. This clearly opens a new window for testing the dark radiation component, since, for example, a smaller value for $\cvis$ could indicate possible non standard interactions (see e.g. \cite{couplings}). A value of $\cvis$ different from zero, as expected in the standard scenario, has been detected in \cite{trotta} and confirmed in subsequent papers
\cite{dopo,darkr}. More recently, a value of the effective sound speed $\ceff$ smaller than the standard value of $1/3$ has been claimed in Ref.~\cite{Smith:2011es}.

From this discussion is clear that current cosmological data analyses 
are sensitive to dark radiation properties and that the latest constraints
on these parameters are showing interesting deviations from their expected standard values. It is therefore timely to investigate the impact of non-standard dark radiation
properties in the determination of cosmological parameters related to different
sectors as dark energy or inflation.

In this paper we derive bounds from the most recent cosmological data on the dark radiation parameters $\neff$, $\ceff$, and $\cvis$ relaxing the usual assumption of a $\Lambda$CDM cosmology, analysing the correlations among the clustering parameters $\ceff$ and $\cvis$ and other key cosmological parameters, as the dark energy equation of state or the scalar spectral index.
We also study these correlations in light of future Planck and COrE CMB data. We shall generate mock CMB data for a cosmology with dark radiation perturbation parameters different from their standard values and then we fit these simulated data via the usual MCMC analysis to an extended non minimal cosmology with standard dark radiation parameters, but varying both the constant and the time varying dark energy equation of state, or the scalar spectral index and its running. 
The paper is organized as follows. Section \ref{sec:i} describes the details of the analysis carried out here, including the cosmological parameters and data sets used in the analyses. In section \ref{sec:ii} the different cosmological scenarios considered here are analyzed and the most important degeneracies among the parameters are carefully explored. Section \ref{sec:iii} summarizes our main conclusions.

\section{Analysis Method}
\label{sec:i}
We perform our analyses considering three different scenarios: we first analyze the Wilkinson Microwave Anisotropy Probe (WMAP) data together with the luminous red galaxy clustering results from the Sloan Digital Sky Survey II (SDSSII)~\cite{beth} and with a prior on the Hubble constant $H_0$ from the Hubble Space Telescope (HST)~\cite{Riess:2009pu}, referring to it as the  ``Baseline model''. Then we add to these data sets SPT data \cite{spt} and we will refer to this case as the ``BaselineSPT model''. We will explore as well the impact of Supernovae Ia (SNIa) luminosity distance measurements from the Union 2 compilation~\cite{Amanullah:2010vv} in constraining the dark radiation parameters, and we will refer to this case in the following as the ``BaselineSPT-SNIa model''.

 We have modified the Boltzmann \texttt{CAMB} code~\cite{camb} incorporating the two extra dark radiation perturbation parameters $\ceff$ and $\cvis$ and we have extracted the cosmological parameters from current data using a Monte Carlo Markov Chain (MCMC) analysis based on the publicly available MCMC package \texttt{CosmoMC}~\cite{Lewis:2002ah}. We sample the following six-dimensional standard parameters: the baryon and cold dark matter densities ($\omega_b\equiv\Omega_bh^{2}$ and $\omega_c\equiv\Omega_ch^{2}$), the ratio between the sound horizon and the angular diameter distance at decoupling $\Theta_{s}$, the optical depth $\tau$, the scalar spectral index $n_s$, and the amplitude of the primordial spectrum  $A_{s}$. We consider purely adiabatic initial conditions and we impose spatial flatness.  We also include the effective number of relativistic degrees of freedom $\neff$, the effective sound speed $\ceff$ and the viscosity parameter $\cvis$. Table \ref{tab:priors} shows the flat priors considered on the different cosmological parameters.
Finally, we generate a mock data set for the ongoing Planck CMB experiment, with $\cvis$ different from its standard value and with $w=-1$ and $n_s=0.96$. Then we fit these mock data using a MCMC analysis to different extensions of the minimal cosmological model in which the dark radiation is standard. The three possible extensions we consider are: \emph{(a)} a $\Lambda$CDM model with a running spectral index $n_{\rm run}$, \emph{(b)} the $w$CDM model in which we include the possibility of a dark energy equation of state parameter $w$ different from $-1$, and \emph{(c)} the $w(a)$CDM model in which we assume an equation of state evolving with redshift. The reconstructed values of the dark energy equation of state and of the running spectral index will be, in general, different from the values used in the mocks and, in the case of the dark energy equation of state $w$, different from the value expected within the $\Lambda$CDM model. We shall also explore the impact of future CMB data from the COrE mission~\cite{core}, performing an equivalent forecast to the one we present here for Planck.

\begin{table}[h!]
\begin{center}
\begin{tabular}{c|c}
\hline\hline
 Parameter & Prior\\
\hline
$\Omega_{b}h^2$ & $0.005 \to 0.1$\\
$\Omega_{c}h^2$ & $0.01 \to 0.99$\\
$\Theta_s$ & $0.5 \to 10$\\
$\tau$ & $0.01 \to 0.8$\\
$n_{s}$ & $0.5 \to 1.5$\\
$\ln{(10^{10} A_{s})}$ & $2.7 \to 4$\\
$\neff$ &  $0 \to 9$\\
$\cvis$ &  $0 \to 1$\\
$\ceff$ &  $0 \to 1$\\
$n_{\rm run}$ & $-0.2\to0.1$\\
$w (w_0)$ & $-2 \to 0$\\
$w_a$ &  $-1 \to  1$\\
\hline\hline
\end{tabular}
\caption{Flat priors for the cosmological parameters considered here.}
\label{tab:priors}
\end{center}
\end{table}

\section{Results}
\label{sec:ii}
\subsection{Baseline models}
We consider a cosmological model described by the following set of parameters:
\begin{equation*}
  \{\omega_b,\: \omega_c,\: \Theta_s,\: \tau,\:w, \: n_s,\: \log[10^{10}A_{s}],\: \neff, \: \cvis, \: \ceff\},
\end{equation*}
Notice from the results in the first column of Tab.~\ref{tab:constraints1} that in the ``Baseline model'', (i.e. the one with WMAP7+SDSSII+$H_0$ data), the preferred value for the effective number of relativistic degrees of freedom is $\neff=5.82^{+0.60}_{-0.84}$, considerably higher than the standard model prediction $\neff=3.04$. The addition of SPT data in the ``BaselineSPT model'' decreases the preferred value of $\neff$, making it closer to its canonical value. Finally, the addition of SNIa data in the ``BaselineSPT-SNIa model'' brings the value of $\neff$ even closer to $3.04$, however, it is still $\sim 2\sigma$ away from the former value. These results seem to agree with the excess of radiation claimed in the literature in previous analyses~\cite{Mangano:2006ur,Hamann:2007pi,Reid:2009nq,Komatsu:2010fb,Hamann:2010pw,spt,darkr,Hamann:2011hu,Nollett:2011aa}. Regarding the dark radiation perturbation parameters $\cvis$ and $\ceff$, their preferred values are close to their standard values of $1/3
 $, being $\ceff$ much better constrained from current data than $\cvis$. 

\begin{figure}[h!]
\includegraphics[scale=0.4]{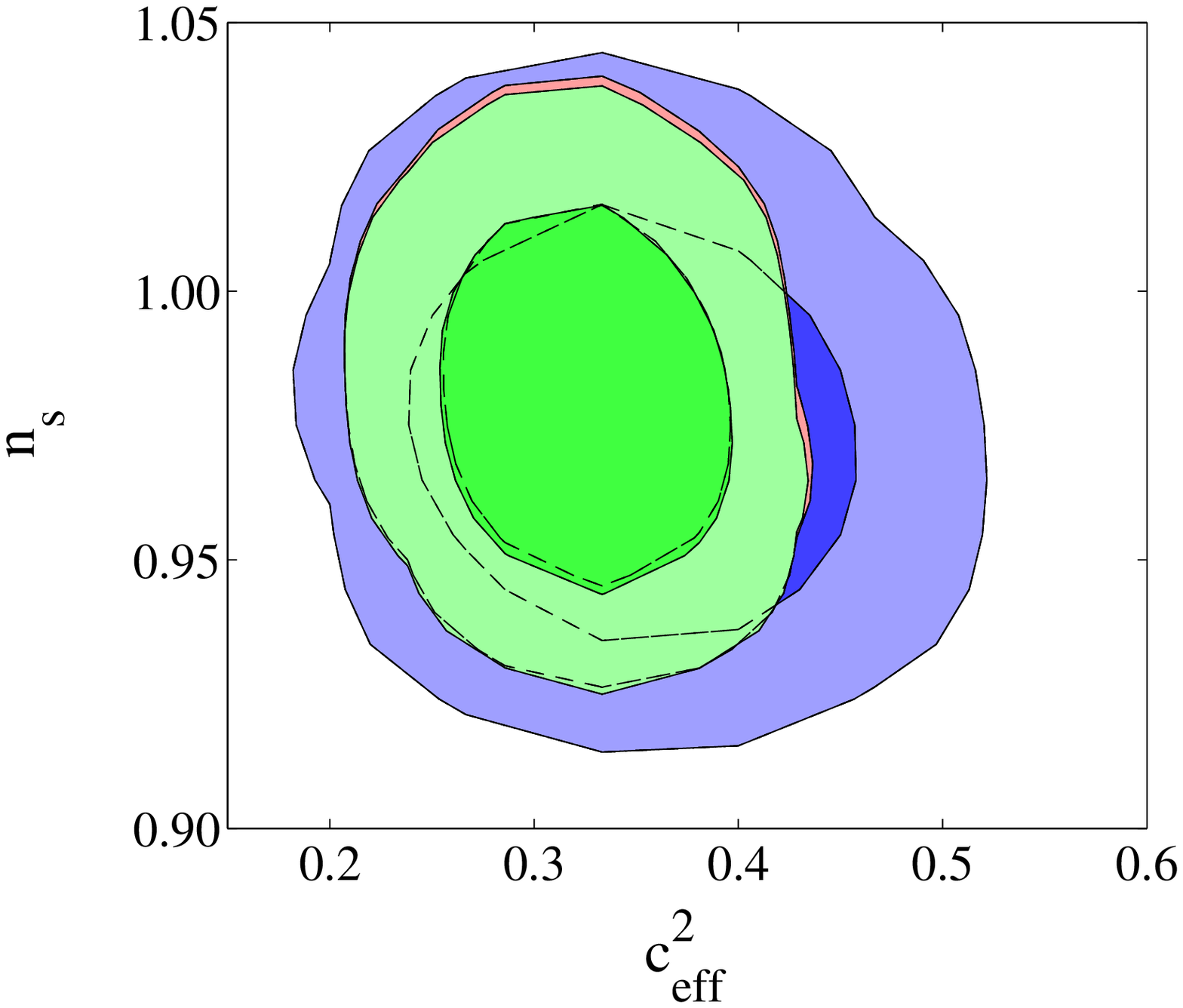}
\includegraphics[scale=0.4]{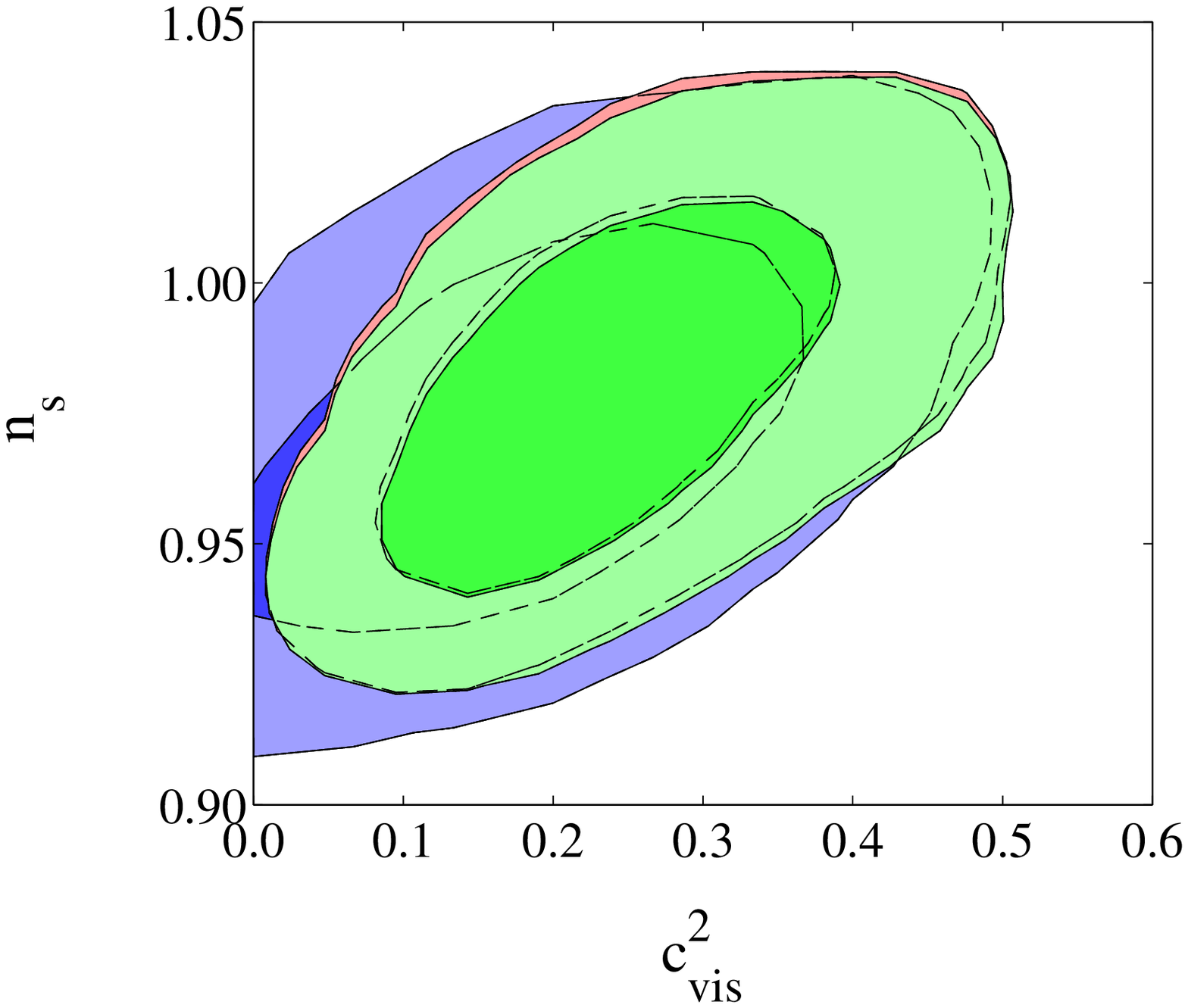}
\includegraphics[scale=0.4]{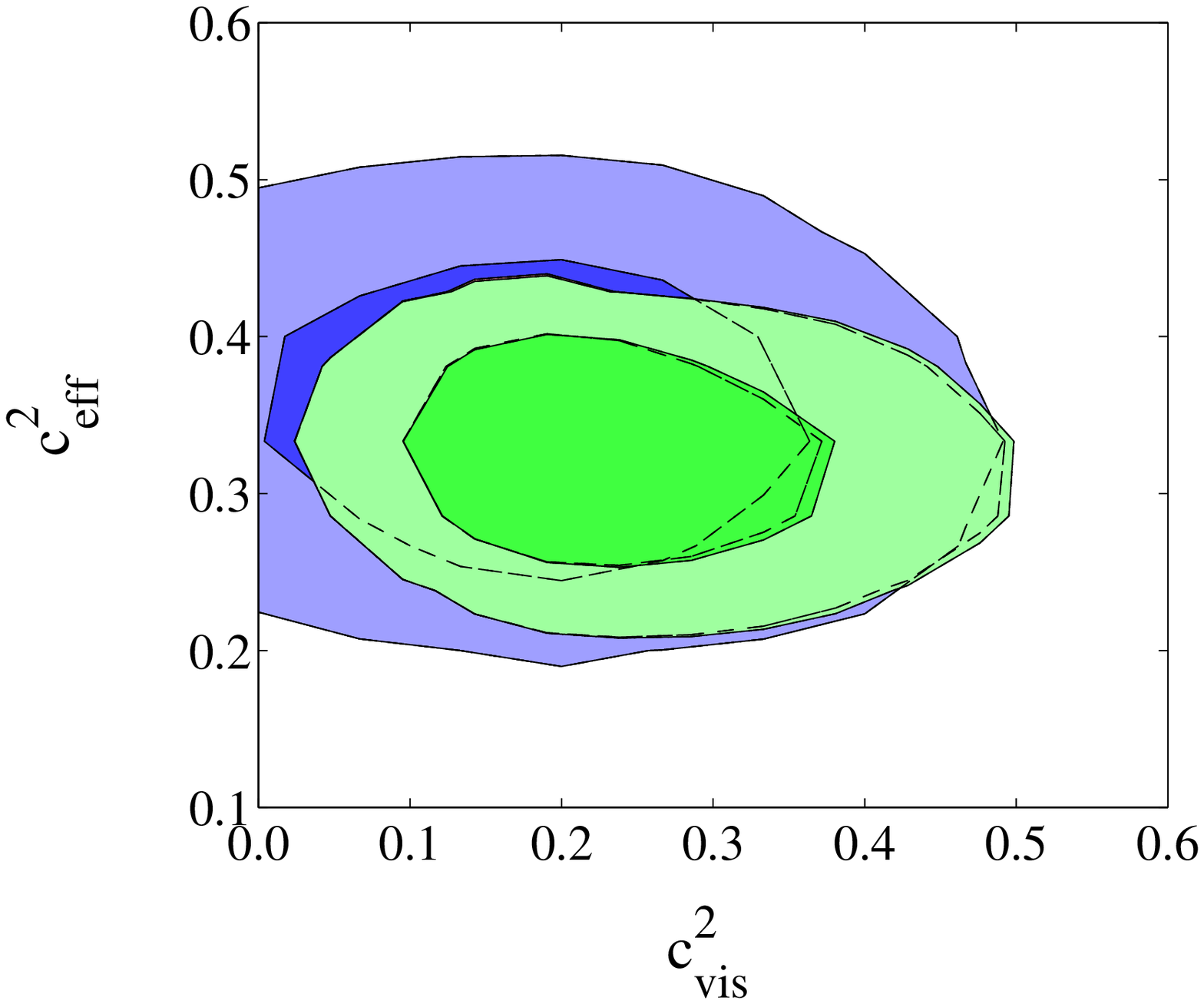}
\caption{The top, middle and bottom panels depict the $68\%$ and $95\%$ CL bounds in the $\ceff-n_s$, $\cvis-n_s$ and $\cvis-\ceff$ planes, respectively. The blue, red and green regions refer to the Baseline, BaselineSPT and BaselineSPT-SNIa models, respectively.} 
\label{fig:ns_ceff_cvis}
\end{figure}

\begin{figure}[h!]
\includegraphics[scale=0.4]{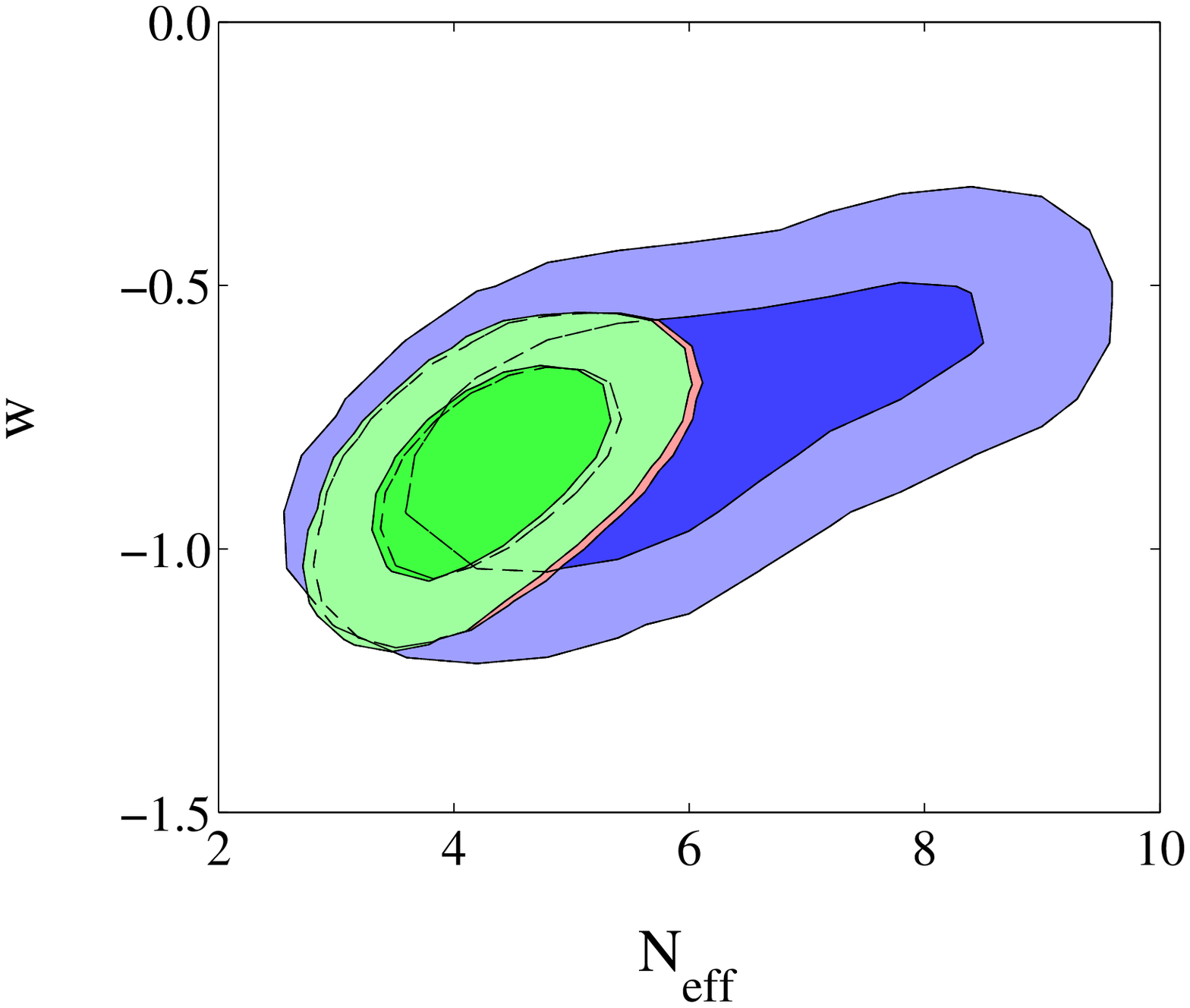}
\includegraphics[scale=0.4]{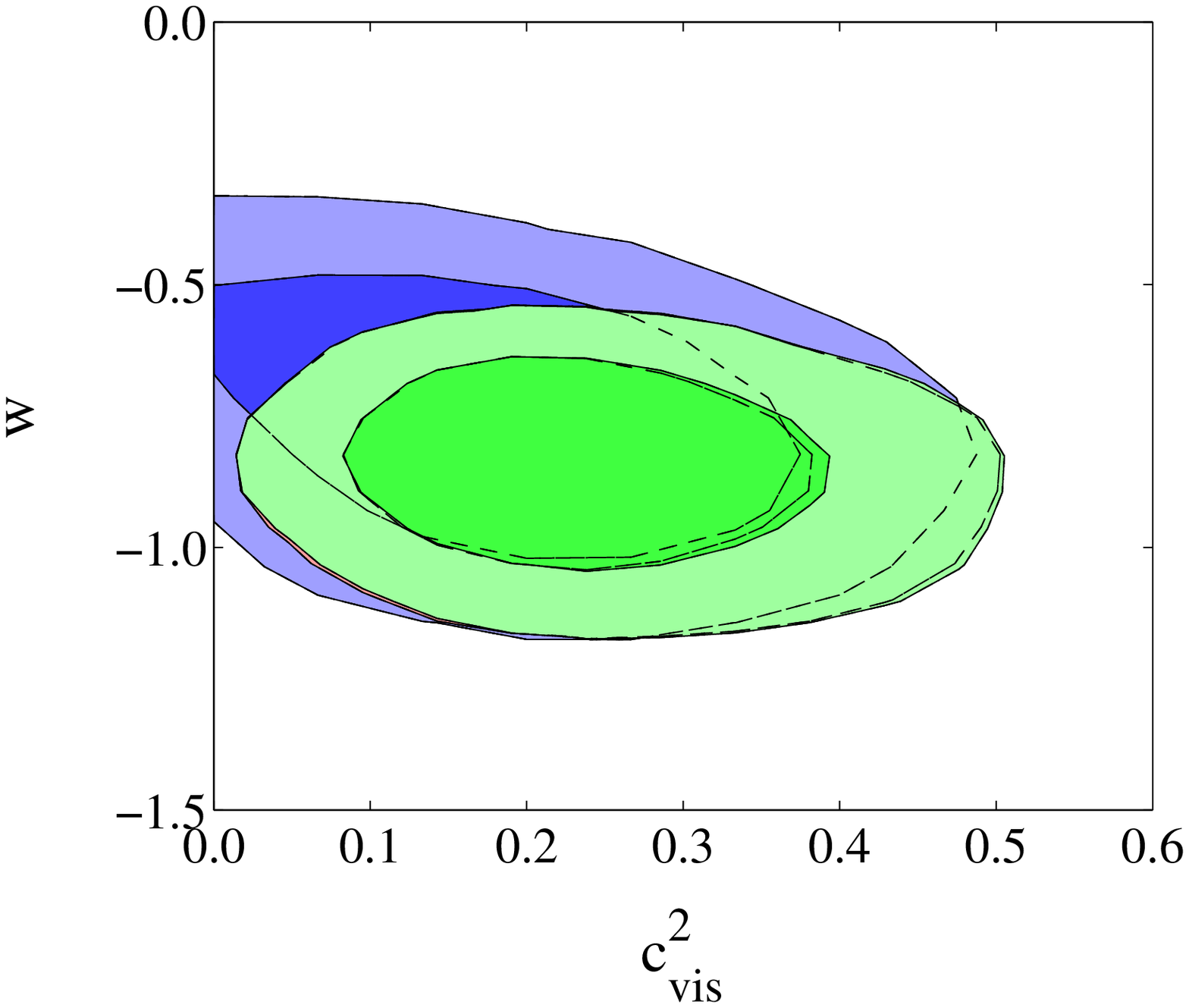}
\caption{The top and bottom panels depict the $68\%$ and $95\%$ CL bounds in the $\neff-w$ and $\cvis-w$ planes, respectively. The blue, red and green regions refer to the Baseline, BaselineSPT and BaselineSPT-SNIa models, respectively.}

\label{fig:w}
\end{figure}

There exists a degeneracy between $\ceff$ and $n_s$ and between $\cvis$ and $n_s$, see the top and middle panels of Fig.~\ref{fig:ns_ceff_cvis}. These degeneracies are related to the fact that if $n_s$ increases (decreases),
the power at low multipoles decreases (increases),
while the power at high $\ell$ increases (decreases) to a lesser extent.
This effect could be compensated by an increase (decrease) in the viscosity parameter $\cvis$, that leads to a decrease (increase) of the power at all scales.
Concerning the degeneracy between $\ceff$ and $n_s$, it is mainly related to the degeneracy between $\ceff$ and $\cvis$ (see the lower panel of Fig.~\ref{fig:ns_ceff_cvis}). There also exist degeneracies between $w$ and $\cvis$ and between $w$ and $\neff$, see Fig.~\ref{fig:w}. As we shall explain in the following section, a value of $w> -1$ shifts the positions of the CMB acoustic peaks to lower multipoles $\ell$; this effect could be compensated by a decrease of $\cvis$ or by an increase of $\neff$. The degeneracy between $w$ and $\cvis$ gets alleviated when information on high $\ell$ multipoles from SPT is considered in the analysis.

\begin{figure}[h!]
\includegraphics[scale=0.4]{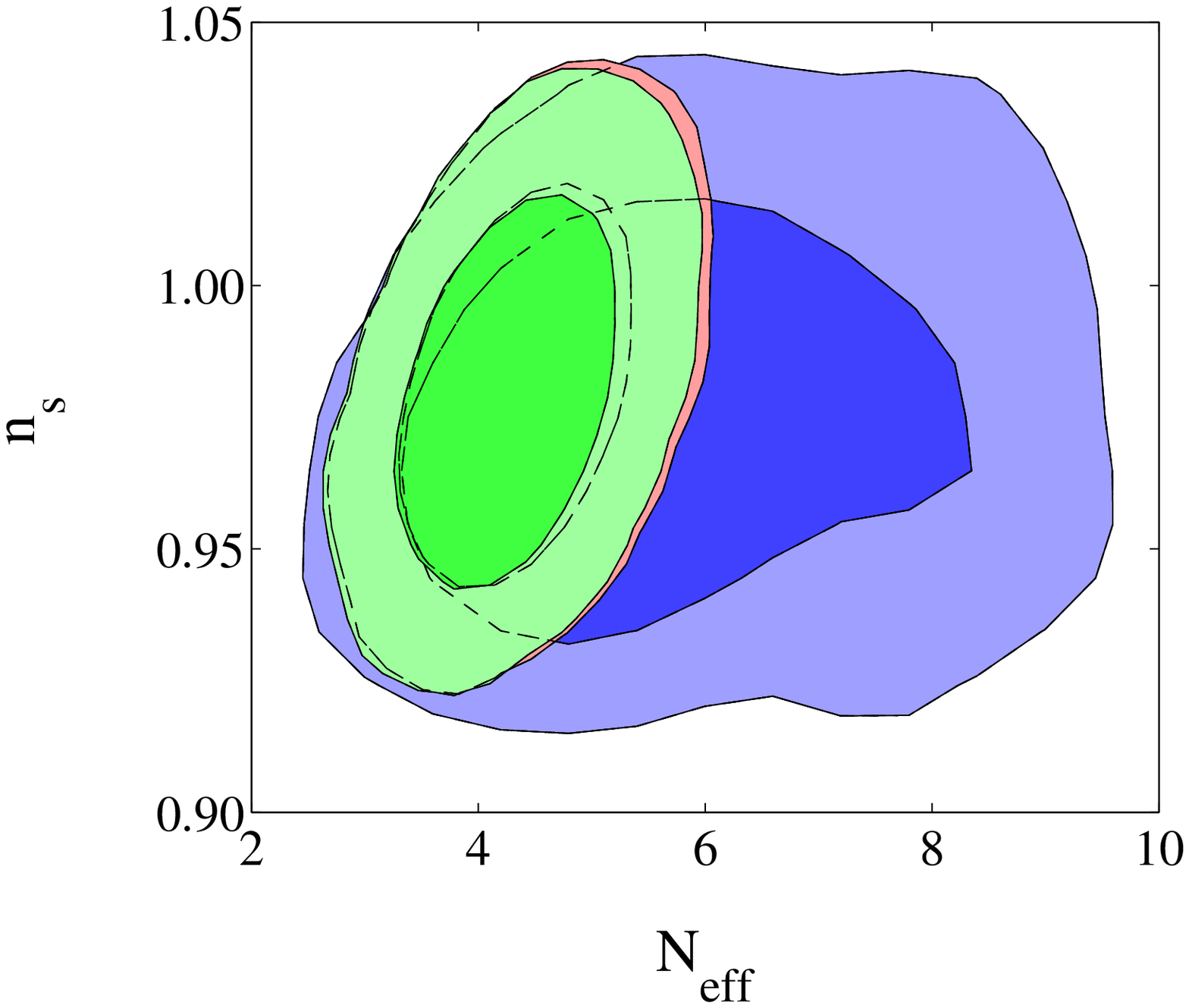}
\includegraphics[scale=0.4]{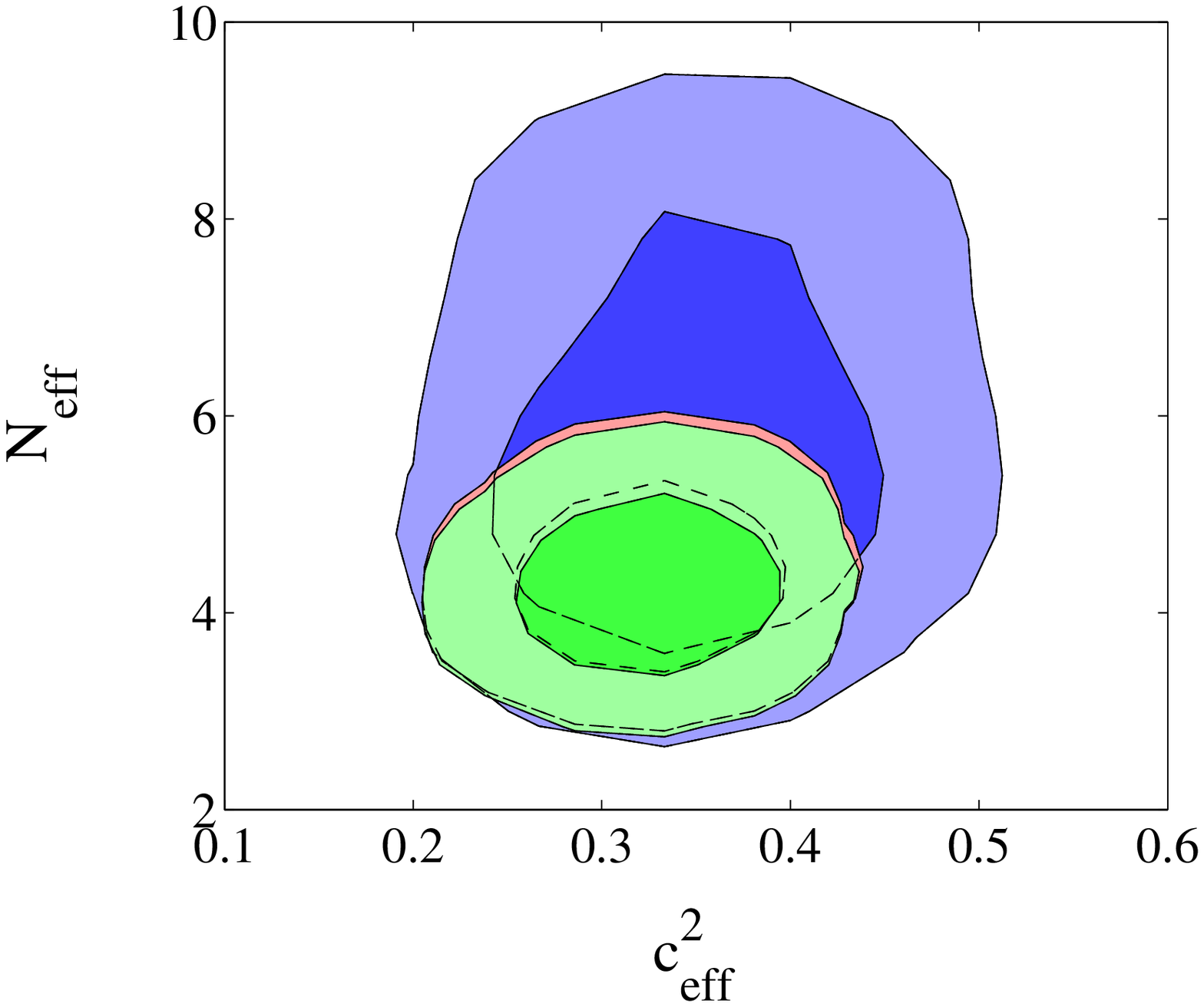}
\caption{The top and the bottom panels depict the 68\%  and 95\% CL bounds in the $\neff-n_s$ and $\ceff-\neff$ planes, respectively. The blue, red and green regions refer to the Baseline, BaselineSPT and BaselineSPT-SNIa models respectively.}

\label{fig:neff}
\end{figure}

A change on the scalar spectral index $n_s$ can also be compensated by a change in $\neff$, see the upper panel of Fig.~\ref{fig:neff}.  This degeneracy only affects the amplitude of the CMB peaks: a higher $\neff$ will reduce the amplitude of the CMB peaks at $\ell>200$ due to a higher Silk damping, which in turn is due to the increased expansion rate~\cite{Hou:2011ec}.

\begin{table*}[th!]\footnotesize
\begin{center}
\begin{tabular}{lcccc}
\hline \hline
& Baseline &  BaselineSPT & BaselineSPT-SNIa \\
& model & model & model\\
\hline
\hspace{1mm}\\
$w$ & $-0.76 \pm 0.15$ & $-0.85 \pm 0.12$ &  $-0.85 \pm 0.12$\\ 
\hspace{1mm}\\
${\neff}$ & $5.82^{+0.60 +2.71}_{-0.84  -2.12}$ & $4.38^{+0.27 +1.07}_{-0.31 -0.98}$ & $4.29^{+0.26 +1.05}_{-0.31 -0.96}$\\
\hspace{1mm}\\
$\cvis$ & $0.21^{+0.10 +0.21}_{-0.10 -0.18}$ & $0.24^{+0.032 +0.17}_{-0.052 -0.13}$  & $0.25^{+0.03 +0.42}_{-0.06 -0.13}$\\
\hspace{1mm}\\
$\ceff$ & $0.35^{+0.01 +0.05}_{-0.02 -0.05}$ & $0.33^{+0.006 +0.024}_{-0.007 -0.024}$ & $0.33^{+0.01 +0.02}_{-0.01 -0.03}$\\
\hspace{1mm} \\
$n_s$ & $0.976 \pm 0.026$ &  $0.982 \pm 0.024 $ & $0.980 \pm 0.024 $\\
\hline
\hline
\end{tabular}
\caption{Constraints on the cosmological parameters for the three Baseline models described  in the text. We report the $68\%$ and $95\%$ CL limits for the dark radiation parameters, and the mean and the standard deviation of the posterior distribution for the other cosmological parameters.}
\label{tab:constraints1}
\end{center}
\end{table*}
\begin{table*}[th!]\footnotesize
\begin{center}
\begin{tabular}{lcccccc}
\hline \hline
& $\Lambda$CDM + $n_{\rm run}$ (Planck) & $\Lambda$CDM + $n_{\rm run}$ (COrE) & $w$CDM (Planck) & $w$CDM (COrE) & $w(a)$CDM (Planck) & $w(a)$CDM (COrE) \\
\hline
\hspace{1mm}\\
$w$ & -1 & -1 & $-0.70 \pm 0.05$ & $-0.63 \pm 0.05$ & --- & ---\\ 
\hspace{1mm}\\
${\neff}$ & 3.04 & 3.04 & 3.04 & 3.04 & 3.04 & 3.04 \\
\hspace{1mm}\\
$n_s$ & $1.002 \pm 0.004$ 
& $0.999 \pm 0.002$ 
& $1.007 \pm 0.004$ 
& $1.004 \pm 0.002$ 
& $1.007 \pm 0.004$ 
& $1.007 \pm 0.002$ 

\\
\hspace{1mm}\\
$n_{\rm run}$ & $-0.035 \pm 0.005$ & $-0.038 \pm 0.003$  
& 0 & 0 & 0 & 0\\
\hspace{1mm}\\
$w_0$ & --- & --- & --- & --- & 
$-1.19 \pm 0.10$ & 
$-0.99 \pm 0.05$\\
\hspace{1mm}\\
$w_a$ & --- & --- & --- & --- & 
$0.77 \pm 0.23$ 
& $0.88\pm 0.06$ \\
\hspace{1mm}\\
\hline
\hline
\end{tabular}
\caption{Constraints on the cosmological parameters for each of the Plank and COrE mock data sets described in the text. We report the mean and the standard deviation of the posterior distribution. We have set $\ceff=1/3$ and $\cvis=0.1$ in the mock data sets used as fiducial models. Then, we have fitted these data to a model with canonical values for the dark radiation perturbation parameters, i.e. $\ceff=1/3$ and $\cvis=1/3$.}
\label{tab:contstraints}
\end{center}
\end{table*}

\subsection{Future Planck and COrE CMB data analysis}
In the following we shall present the reconstructed values of $n_s$, $n_{\rm run}$, $w$, $w_0$ and $w_a$ which will result from a fit of Planck and COrE mock data (generated with non standard values for the dark radiation perturbation parameters, $\cvis=0.1$)  to a cosmology with a standard value for the dark radiation parameter $\cvis=1/3$ but with a running spectral index or a time varying dark energy component. We do not consider here $\ceff\neq 1/3$ due to the tighter current bounds on this parameter, when compared to the current constraints on $\cvis$.

\subsubsection{$\Lambda$CDM + $n_{\rm run}$}

For this scenario we consider the following set of parameters:
\begin{equation*}
  \{\omega_b,\: \omega_c,\: \Theta_s,\: \tau,\: n_s,\: \log[10^{10}A_{s}],\: n_{\rm run}\}~.
\end{equation*}
In general, the spectrum of the scalar perturbations is not exactly a power law but it varies with scale. Therefore one must consider the scale dependent running of the spectral index $n_{\rm run}\:=\:dn_s/d\ln k$.
Following \cite{Kosowsky:1995aa}, the power spectrum for the scalar perturbations reads
\begin{equation*}
P(k) \equiv A_sk^{n(k)} \propto \left( k\over k_0 \right)^{n_s\, +\, \ln(k/k_0)
(dn/d\ln k)\, +\, \cdots }~,
\end{equation*}
being $k_0=0.05$~Mpc$^{-1}$ the pivot scale.
\begin{figure}[h!]
\includegraphics[scale=0.4]{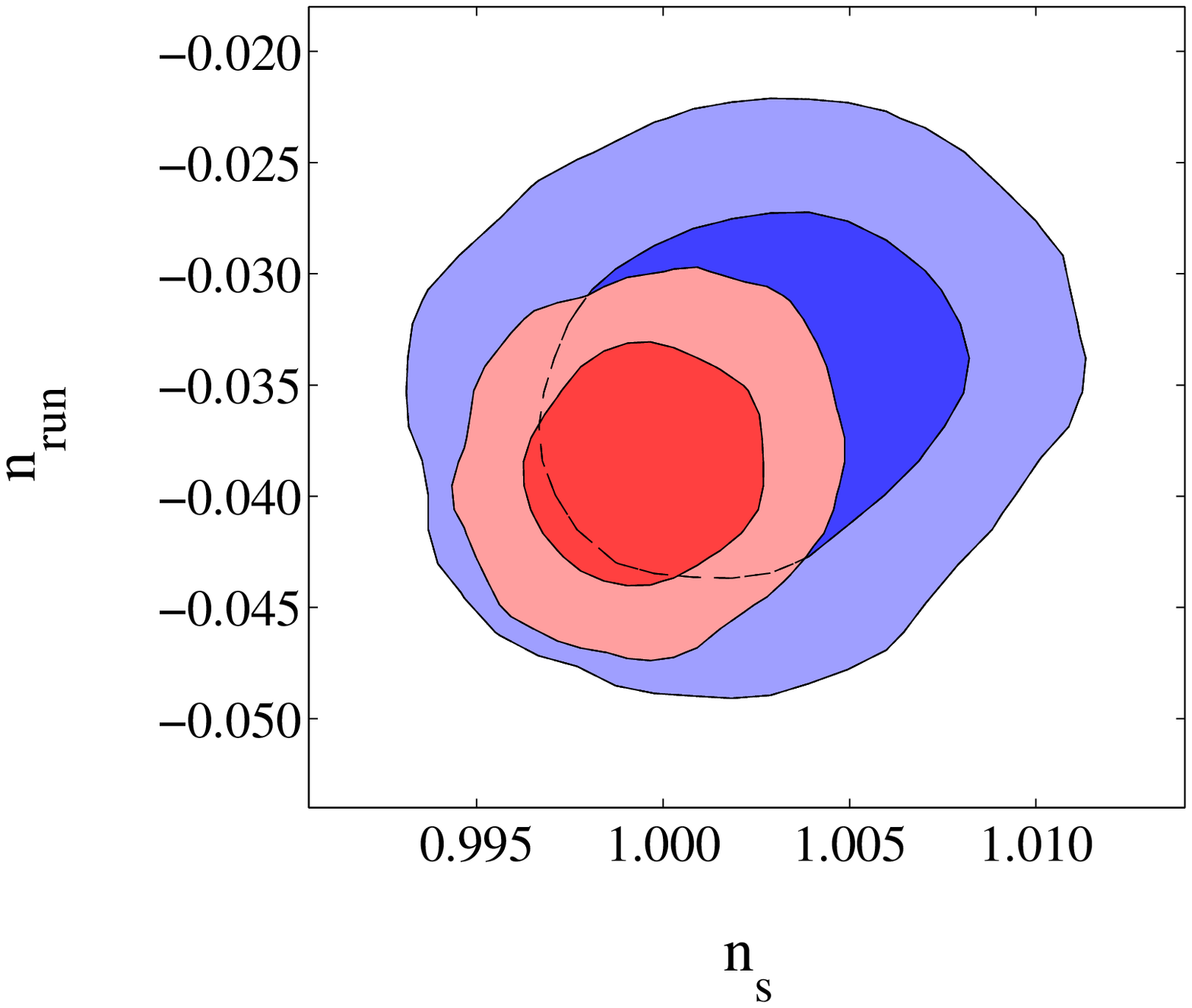}
\caption{68\% and 95\% CL allowed regions in the $n_s-n_{\rm run}$ plane
from MCMC fits of Planck (blue regions) and COrE (red regions) CMB mock data.}
\label{fig:ns_nrun}
\end{figure}
The correlation between $n_s$ and $n_{\rm run}$ is shown in Fig.~\ref{fig:ns_nrun}.
As stated in \cite{Hannestad:2001nu}, the parameter that is constrained by cosmological data  is the effective spectral index $n^\prime\:=\:n_s+\ln (k/k_0) (dn/d\ln k)$. This is the reason for the circular allowed regions in the $n_s-n_{\rm run}$ plane.
 The first and second columns of Tab.~\ref{tab:contstraints} show that, if a cosmology with $n_{\rm run}=0$ but with non standard dark radiation perturbation parameters ($\cvis=0.1$) is fitted to a cosmology with standard dark radiation parameters but with $n_{\rm run}\neq 0$,  the reconstructed value of the running spectral index will differ from zero at a high statistical significance.

Finally, for the case of the simulated cosmology here with $\cvis < 1/3$, the reconstructed value of $n_s$ is consistent with a Harrison-Zel'dovich scale invariant primordial power spectrum within one sigma. Setting the properties of dark radiation is therefore mandatory since it is highly correlated with the spectral index of the spectrum of primordial perturbations, key to distinguish among the different inflationary models.

\subsection{$w$CDM}

Here we consider a cosmological model including a dark energy fluid characterized by a constant equation of state $w$ as a free parameter. We consider the following set of parameters:
\begin{equation*}
\{\omega_b,\: \omega_c,\: \Theta_s,\: \tau,\: n_s,\: \log[10^{10}A_{s}],\: w\}~.
\end{equation*}
As stated in a previous work \cite{us}, there exists a degeneracy between the number of the extra dark radiation species and the dark energy equation of state, see Fig.~\ref{fig:w} (upper panel). One of the main effects of a $\neff>3.04$ comes from the change of the epoch of the radiation matter equality, and consequently, from the shift of the CMB acoustic peaks, see Ref.~\cite{Hou:2011ec} for a detailed study. The position of the acoustic peaks is given by the so-called acoustic scale $\theta_A$, which reads
\begin{equation*}
\theta_A=\frac{r_s(z_{rec})}{r_\theta(z_{rec})},
\end{equation*}
where $r_\theta (z_{rec})$ and $r_s(z_{rec})$ are the comoving angular diameter distance to the last scattering surface and the sound horizon at the recombination epoch $z_{rec}$, respectively. Although 
$r_\theta (z_{rec})$ almost remains the same for different values of $\neff$, $r_s(z_{rec})$ becomes smaller when $\neff$ is increased. Thus the positions of the acoustic peaks are shifted to higher multipoles (smaller angular scales) by increasing the value of $\neff$~\cite{arXiv:0803.0889}. A dark energy component with $w> -1$ will decrease the comoving angular diameter distance to the last scattering surface $r_\theta (z_{rec})$, shifting the positions of the CMB acoustic peaks to larger angular scales, i.e. to lower multipoles $\ell$, compensating, therefore, the effect induced by an increase of $\neff$.
The reconstructed MCMC values for $w$ (see the third and fourth columns of Tab.~\ref{tab:contstraints}) are larger than the value used in the input cosmology $w=-1$, excluding the $\Lambda$CDM scenario with high significance. A dark radiation component which deviates from its standard behavior could therefore be confused with the presence of a dark energy fluid with $w\neq -1$.

\begin{figure}[h!]
\includegraphics[scale=0.4]{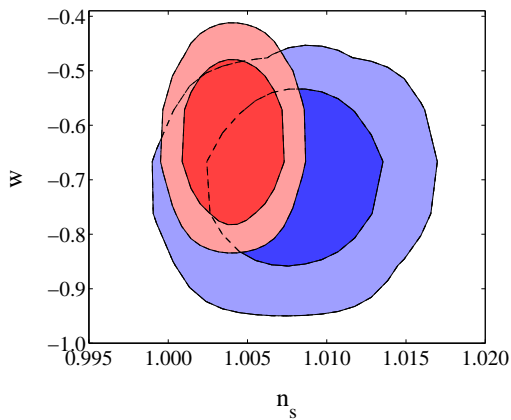}
\caption{Same as Fig.~\ref{fig:ns_nrun} but in the $w-n_s$ plane.} 
\label{fig:w_ns}
\end{figure}

\subsection{$w(a)$CDM model}

\begin{figure}[h!]
\includegraphics[scale=0.4]{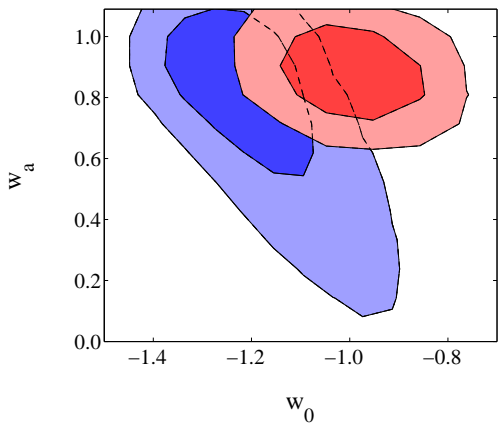}
\caption{Same as Fig.~\ref{fig:ns_nrun} but in the $w_0-w_a$ plane.} 
\label{fig:wa}
\end{figure}

We also consider models of the dark energy in which the equation of state of the dark energy component varies with time. We use a parameterization that has been extensively explored in the literature~\cite{Chevallier:2000qy,Linder:2002et,Albrecht:2006um,Linder:2006sv}: 
\begin{equation*}
w(a)=w_0+w_a(1-a)~,
\end{equation*}
where $w_0$ is the equation of state parameter at present, while $w_a = -2 {dw}/{d\ln{a}}|_{a=1/2}$~\cite{Linder:2002et,Linder:2004ng}. 
We consider the following set of parameters:
\begin{equation*}
\{\omega_b,\omega_c, \Theta_s, \tau, n_s, \log[10^{10}A_{s}], w_0, w_a\}~.
\end{equation*}
The fifth and sixth columns of Tab.~\ref{tab:contstraints} show the reconstructed values of $w_0$ and $w_a$ after fitting the Plank and COrE mock data generated with a non standard viscosity parameter $\cvis=0.1$ but with $w=-1$ to a cosmology with standard dark radiation but with the time varying dark energy equation of state $w(a)$ used here. The correlation between $w_0$ and $w_a$ is shown in Fig.~\ref{fig:wa}. The reconstructed values that we find for Planck (COrE) mock data are $w_0= -1.19 \pm 0.10$ and $w_a= 0.77 \pm 0.23$ ($w_0= -0.99 \pm 0.05$ and $w_a=0.88\pm 0.06$) at 68 $\%$ CL, values which are consistent with the current constraints on these two dark energy parameters, see Ref.~\cite{Komatsu:2010fb}. Therefore it is crucial to unravel the nature of the dark radiation component since if it turns out to be non standard, future cosmological data might be misinterpreted as a time varying dark energy fluid. 

\section{Conclusions}
\label{sec:iii}
Dark radiation in the Standard Model of elementary particles is made of three light neutrinos. However, many extensions of the Standard Model predict an extra dark radiation component parameterized in terms of the relativistic degrees of freedom. Such is the case of sterile neutrinos, axions or other light degrees of freedom produced along the thermal history of the universe. This extra dark radiation component will be characterized not only by its abundance but also by its clustering properties, as its effective sound speed and its viscosity parameter. If dark radiation is made of sterile neutrinos it should have an effective sound speed $\ceff$ and a viscosity parameter $\cvis$ such that $\ceff=\cvis=1/3$. However, other relativistic species might not behave as neutrinos, being $\cvis$ and $\ceff$ different from their canonical values. Current bounds on the number of relativistic species and on the dark radiation perturbation parameters $\ceff$ and $\cvis$ have been computed using up to date cosmological data. We find a strong degeneracy between $\cvis$ and the scalar spectral index of primordial perturbations, as well as between $\cvis$ and the dark energy equation of state $w$. The last degeneracy is alleviated when CMB data from the SPT experiment is added in the MCMC analyses. A question which naturally arises from the presence of these degeneracies is whether or not future CMB data will be able to distinguish among different dark radiation scenarios. We have generated mock CMB data for the ongoing Planck experiment and the future COrE mission with non standard values for the dark energy perturbation parameters. Then, we have fitted these data to a canonical dark radiation scenario with $\cvis=\ceff=1/3$ but with a running spectral index or with a dark energy component with $w\neq -1$, finding that non standard values for the dark radiation perturbation parameters may be misinterpreted as a scale invariant power spectrum of primordial fluctuations or as cosmologies with a running spectral index or a time varying dark energy component with high significance. 

\vspace {1cm}
\begin{acknowledgments}
O.M. is supported by the Consolider Ingenio project CSD2007-00060, by PROMETEO/2009/116, by the Spanish Ministry Science project FPA2011-29678 and by the ITN Invisibles PITN-GA-2011-289442. 
This work is supported by PRIN-INAF, "Astronomy probes fundamental physics".
Support was given by the Italian Space Agency through the ASI contracts Euclid- IC (I/031/10/0).

\end{acknowledgments}



\end{document}